\documentclass[referee]{cjaa}
\usepackage{graphicx}                   
\input{epsf.sty}                        
\input{psfig.sty}                       

\begin{document}

\title{Algorithm of ensemble pulsar time}

\volnopage{Vol.0 (200x) No.0, 000--000}      
\setcounter{page}{1}          

\author{Alexander E. Rodin \inst{1}}
\offprints{A.E.Rodin}
\institute{Pushchino Radio Astronomical Observatory, Pushchino, Moscow region,
   142290, Russia
   \email{rodin@prao.psn.ru}
   }

\date{Received~~2005 month day; accepted~~2005~~month day}

\abstract{ An algorithm of the ensemble pulsar time based on the
Wiener filtration method has been constructed. This algorithm has
allowed the separation of the contributions of an atomic clock and a
pulsar
itself to the post-fit pulsar timing residuals. The method has been
applied to the timing data of the millisecond pulsars PSR B1855+09 and
PSR B1937+21 and allowed to filter out the atomic scale component from
the pulsar phase variations. Direct comparison of the terrestrial time
TT(BIPM96) and the ensemble pulsar time PT$_{\rm ens}$ has displayed
that the difference TT(BIPM96) -- PT$_{\rm ens}$ is within
$\pm0.4\;\mu$s range. A new limit of gravitational wave background
based on the difference TT(BIPM96) -- PT$_{\rm ens}$ was established
to be $\Omega_g {\rm h}^2\sim 10^{-10}$.  
}

\authorrunning{A.E.Rodin}
\titlerunning{Algorithm of ensemble pulsar time}

\maketitle

\section{Introduction}

The discovery of pulsars in 1967 (\cite{Hewish68}) and millisecond
pulsars in 1982 (\cite{Backer82}) and consequent observations had
shown clearly that their rotational stability allowed them to be used
as astronomical clocks.

In this paper the author presents a method of forming the ensemble
pulsar time scale (PT). The method is based on the optimal Wiener
filter.  In Sect.~\ref{sect2} principles of pulsar timing are
described with regard to time scales. Sect.~\ref{sect3} contains a
theoretical algorithm of the Wiener filter and construction of the
ensemble pulsar time scale. Sect.~\ref{sect4} presents results,
Sect.~\ref{sect5} discusses an application of the algorithm to timing
data of pulsars PSR B1855+09 and PSR B1937+21 (\cite{kaspi94}).

\section{Pulsar timing}\label{sect2}

An observer which is situated on the Earth rotating around its axis
and moving around the Sun receives with a radio telescope a pulsar
signal during an integration time to obtain sufficient signal-to-noise
ratio.  Time of arrival (TOA) of the integrated pulses are measured
with the observatory freguency standard (e.g. H-maser) by the maximum
of the cross-correlation between the integrated pulse and the pulse
template. The obtained topocentric TOAs $\tau_N$ are in the scale of
the local frequency standard and therefore required to be converted to
the barycentric time scale via the following expressions
(\cite{mt77}; \cite{oleg90}):
\begin{equation}
UTC=\tau_N+\Delta\tau_1,\; \label{eq1}
TAI=UTC+k,\; 	
TT=TAI+32.184\;{\rm s}, 
\end{equation}
where $\Delta\tau_1$ is the correction of the local scale to the universal
coordinated time UTC. The international atomic time (TAI) differs from
UTC by $k$ integer number of  seconds introduced to reconcile the
lengths of day measured by an atomic clock and the Earth rotation.
TAI is related with the terrestrial time TT by the eq.~(\ref{eq1}),
where the constant shift 32.184 sec prevents a jump between ephemeris
and atomic time. Since a second of TT has various lengths
depending on the position and velocity of the Earth in its orbit then
a transformation from TT to TB scale is required which is performed on
the basis of the paper by Fairhead \& Bretagnon (\cite{fairhead90}).
Once converted from TT to TB the topocentric TOAs need to be
reduced to the barycentre of the Solar system (SSB) according to the
following transformation formula (\cite{mt77}; \cite{oleg90}):
\begin{equation}
T=t-t_0 + \Delta_R(\alpha,\delta,\mu_\alpha, \mu_\delta, \pi)
-\Delta_R(x,e,P_b,T_0,\omega) -D/f^2 +\Delta_{\rm rel} +\Delta t_{\rm
  clock},
\end{equation}
where $t_0$ is the reference epoch, $t$ is the pulsar topocentric TOA in TB
scale, $T$ is TOA at the Solar system barycentre in TB scale,
$\Delta_R(\alpha,\delta,\mu_\alpha, \mu_\delta, \pi)$ is Roemer delay
along the Earth orbit, $\Delta_R(x,e,P_b,T_0,\omega)$ is Roemer delay
along the pulsar orbit, $D/f^2$ is dispersive delay for propagation at the
frequency $f$ (corrected for the Doppler shift) through the interstellar
medium, $\Delta_{\rm rel}$ is time corrections due to relativistic
effects in the Solar system and the pulsar orbit, $\Delta t_{\rm clock}$ is
the offset between the observatory frequency standard and the terrestrial time.

Time of arrivals at the SSB are then used for calculation of the
pulsar  rotational phase (in cycles)
\begin{eqnarray}\label{phase}
N(T)=N_0+\nu T +\frac12\dot\nu{T}^2+\varepsilon(T),
\end{eqnarray}
where $N_0$ is the initial phase at epoch $T=0$, $\nu$, $\dot\nu$ are the
pulsar spin frequency and its derivative respectively at epoch $T=0$,
$\varepsilon(T)$ is phase variations (timing noise). The fitting
procedure includes adjustment of the parameters $N_0$, $\nu$,
$\dot\nu$, $\alpha$, $\delta$ and so on, to minimise the weighted sum
of squared differences between $N(T)$ and the nearest integer. Usually
the pulsar rotational phase residuals are expressed in units of time
$\delta t=\delta N/\nu$. In this paper we deal with the only part of
the residuals that includes the variations of the clock offset $\Delta
t_{clock}(T)$.

When comparing different realisations of atomic time scales between
each other one can see that they are dominated by flicker frequency
noise on intervals of a few months and random walk in frequency on
intervals of years (\cite{Guinot88}), i.e. in the frequency domain the
clock variations have power spectrum of form $1/\omega^{n}$, in the
time domain
clock variations can be expressed in the polynomial form
\begin{equation}
\Delta {t}_{\rm clock}({\cal T})=c_0 + c{\cal T}+\frac 12\dot c{\cal T}^2+ \frac 16\ddot c{\cal T}^3+\ldots.
\end{equation}
One can see that the appearance $\Delta t_{\rm clock}$ in
eq.(\ref{phase}) results in a coupling between pulsar and clock
parameters:
\begin{equation}
N({\cal T})={N}_0'+(1+c)f {\cal T}+\frac 12(f\dot c+(1+c)^2\dot f){\cal T}^2,
\end{equation}
where $\cal T$ is the ideal time scale, $f,\,\dot f$ are the pulsar
frequency and its derivative not subjected to the influence of the
clock parameters. For this reason one should use TOAs expressed in the
best available time scale TT (\cite{GuinotPetit91}).


\section{Filtering technique}\label{sect3}

Let us consider $n$ measurements of a random value ${\bf
r}=(r_1,r_2,\ldots,r_n)$ are given. ${\bf r}$ is a sum of two
uncorrelated values ${\bf r}={\bf s}+\varepsilon$, where ${\bf s}$ is a
random signal to be estimated and associated with the clock contribution,
$\varepsilon$ is random errors associated with the fluctuations of pulsar
rotation. Both values ${\bf s}$ and $\varepsilon$ should be related to the
{\it ideal} time scale since a pulsar on the sky "does not know" about the
time scales used for their timing. The problem of the Wiener
filtration is concluded in estimation of the signal ${\bf s}$ if the
measurements ${\bf r}$ and the covariances $\left<s_i,s_j\right>$ and
$\left<\varepsilon_i, \varepsilon_j\right>,\;(i,j=1,2,\ldots,n)$ are
given (\cite{gubanov97}; \cite{vaseghi2000}). The optimal estimation
of the signal ${\bf\hat s}$ is expressed by the formula
(\cite{gubanov97}; \cite{vaseghi2000})
\begin{equation}\label{signal}
{\bf\hat s}={\bf Q}_{sr}{\bf Q}_{rr}^{-1}{\bf r}
={\bf Q}_{ss}{\bf Q}_{rr}^{-1}{\bf r}
={\bf Q}_{ss}({\bf Q}_{ss}+{\bf Q}_{\varepsilon \varepsilon})^{-1}{\bf r},\\
\end{equation}
where ${\bf Q}_{rr}$, ${\bf Q}_{ss}$ are the covariance matrices of
the noise data ${\bf r}$ and signal ${\bf s}$ respectively, ${\bf
Q}_{sr}$, ${\bf Q}_{rs}$ are the cross-covariance matrices between
${\bf r}$ and ${\bf s}$. The covariance matrix ${\bf Q}_{ss}$ is
calculated as cross-covariances
$\left<{}^k r_i,{}^l r_j\right> = \left<s_i,s_j\right>,\; (k,l=1,2,\ldots,M;\;i,j=1,2,\ldots,n)$, $M$ is a total number of pulsars.
In the formula (\ref{signal}) the matrix ${\bf Q}_{rr}^{-1}$ serves as the
whitening filter. The matrix ${\bf Q}_{ss}$ forms the signal from the whitened data.

The ensemble signal (time scale) is expressed as following
\begin{equation}
{\bf\hat s_{\rm ens}} = \frac{2}{M(M-1)}
\sum_{k=1}^{{M(M-1)\over 2}}{}^k{\bf Q}_{ss}\cdot
\sum_{i=1}^{M}{}{}w_i{}^i{\bf Q}_{rr}^{-1} \cdot{}^i{\bf r},
\end{equation}
where $w_i$ is the relative weight of the $i$th pulsar, $w_i\propto 1/\sigma_i$, $\sigma_i$ is the root-mean-square of the expression 
${}^i{\bf Q}_{rr}^{-1} \cdot{}^i{\bf r}$.

\section{Results}\label{sect4}

The method described in the previous section has been applied to the pulsar
timing data of PSR B1855+09 and B1937+21 (\cite{kaspi94}). Though these data are regular they are unevenly spaced, therefore a cubic spline approximation has been applied to make them uniform with sampling interval 10 days. Such a procedure perturbs a high-frequency component of the data and leaves unchanged a low-frequency component which is of interest.

The common part of the residuals for both pulsars (251 TOAs) has been
taken within the interval $MJD=46450\div 48950$. Since the residuals
after the procedure of dropping their ends have the different mean and
the slope they have been quadratically refitted for consistency with
the classical timing fit. The residuals after all treatments described
above are shown in Fig.~1.

\begin{figure*}\label{res12}
\centering \psfig{width=10cm,figure=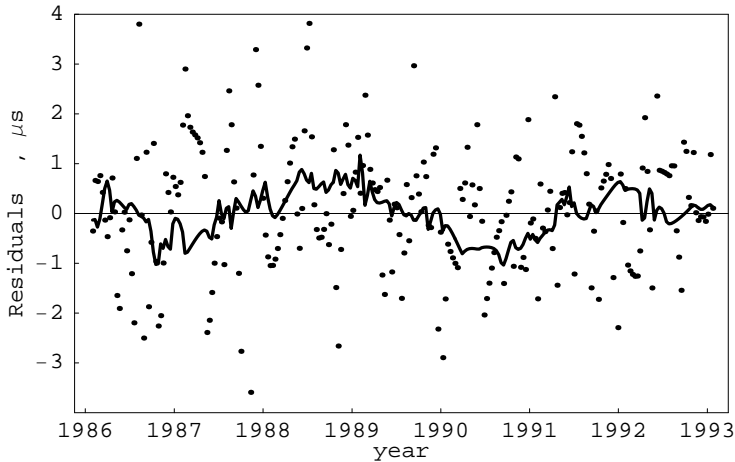}
\caption{The barycentric timing residuals of pulsars PSR B1855+09 (dots)
and PSR B1937+21 (continuous line).}
\end{figure*}

According to (\cite{kaspi94}) the timing data of PSRs B1855+09 and
B1937+21 are in UTC  time scale, hence the signal to be
estimated is the difference UTC -- PT. Fig.~2 shows the signal
estimates based on the residual TOAs of pulsars PSR B1855+09 and PSR
B1937+21. The ensemble signal and the difference UTC -- TT display similar behaviour.

\begin{figure*}\label{signal12}
\centering \psfig{width=10cm,figure=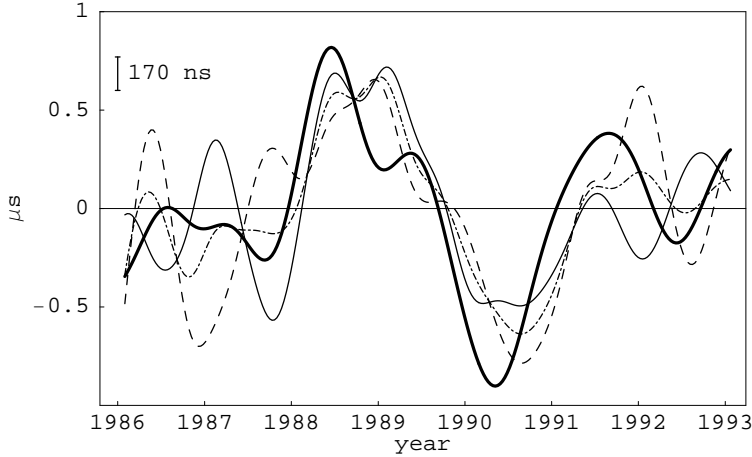}
\caption{Combined clock variations of UTC -- PT in interval
$MJD=46450\div48950$ estimated using the optimal filtering method based on the timing residuals of pulsars PSR B1855+09 (thin line), PSR B1937+21
(dashed line), ensemble UTC -- PT$_{\rm ens}$ (dot-dashed line) and UTC -- TT (solid line).}
\end{figure*}

\section{Discussion}\label{sect5}

For calculation of the fractional instability of a pulsar as a clock a statistic $\sigma_z$ has been proposed (\cite{taylor91}). A detailed numerical algorithm for calculation of $\sigma_z$ has been described in the paper (\cite{matsakis97}).

Fig.~3 presents the fractional instability of PSR B1855+09, PSR
B1937+21 and TT -- PT$_{\rm ens}$. The theoretical lines of $\sigma_z$
(\cite{kaspi94}) in the case when the timing residuals are dominated
by the gravitational wave background with $\Omega_g {\rm h}^2=10^{-9}$
and $10^{-10}$ are plotted in the lower right hand corner. One can see
that $\sigma_z$ of TT -- PT$_{\rm ens}$ crosses line $\Omega_g {\rm
h}^2=10^{-9}$ and approaches $\Omega_g {\rm h}^2=10^{-10}$.

The fractional instability of the TT relative to PT$_{\rm ens}$ is at
level of $10^{-15}$ at 7 years interval and almost one order better
than the fractional instability of the pulsars PSR B1855+09 and PSR
B1937+21. It is expected that reliability of TT -- PT$_{\rm ens}$
estimation will grow up by increasing the number of pulsars
participating in PT$_{\rm ens}$ as $M(M-1)/2$ (the number of
cross-correlations). Currently the accuracy of the filtering method
described above without contribution of the uncertainty of TT algorithm is
estimated at level 170 ns. This uncertainty is obtained as
root mean square value of the data points taken within the smoothing
interval of span $m$. The span $m$ was calculated from the
equivalent bandwidth of the low-pass filter applied to the ensemble
data for more easy comparison with UTC -- TT line.
The uncertainty of PT$_{\rm ens}$ may, in principle,
reach the nanosecond level if to use all the observed highly-stable
millisecond pulsars.
\begin{figure*}\label{sigma-z}
\centering \psfig{width=10cm,figure=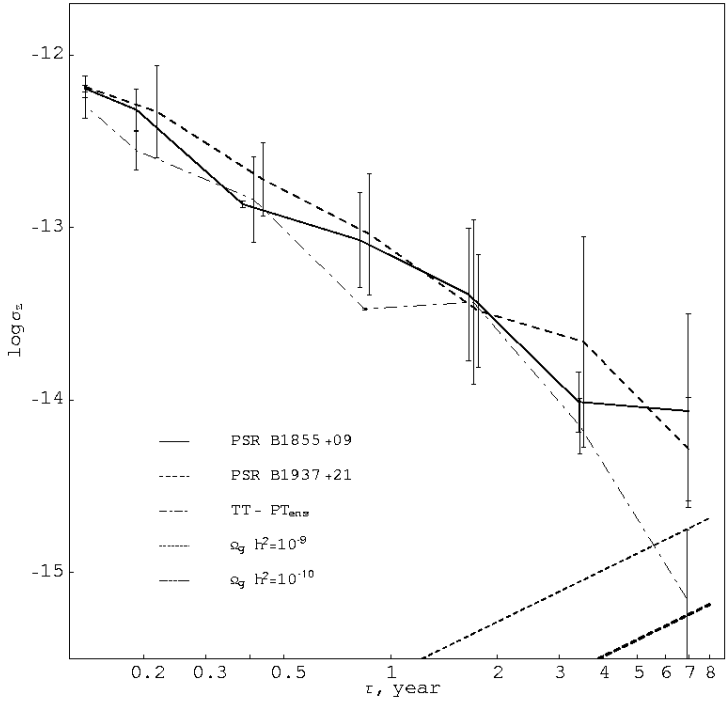}
\caption{The fractional instability $\sigma_z$ for pulsars PSR B1855+09
(solid line), PSR B1937+21 (dashed line) and TT -- PT$_{\rm ens}$
(dot-dashed line).}
\end{figure*}

The method proposed can not distinguish the 2nd order polynomial
trends in the reference clock and the pulsar phase due to pulsar
period slowing-down. However, this is not a problem if to consider the
timing data at more long intervals and process them off-line. Under
such processing the long-term details appears as the data span is
increased.

The low fractional accuracy of $\dot P$ mentioned in the paper
(\cite{GuinotPetit91}) produces no disadvantages when processing off-line
since no prediction of the pulsar rotational phase is performed. However if one does need to predict a behaviour of the concrete atomic scale variations,
e.g. UTC, then this can be done on the basis of the UTC -- PT$_{\rm ens}$ data by using standard forecasting methods for the time series, e.g. the auto-regression method with reservation that only relatively short-term variations without quadratic trend are forecasted. Under such an approach the unsatisfactory fractional accuracy of the spin period derivative does not play significant role since the phase variations are predicted rather than an absolute value.

The proposed filtering method can be applied in "inverse" form: one pulsar
and a few reference clocks. In such case it is also possible to separate
the pulsar timing noise and the clock variations relative to the ideal time
scale rather than to obtain a simple clock difference.



\end{document}